# On the construction of a local curvilinear coordinate system conforming to the native curved geometry of the plasma focus sheath

S K H Auluck
International Scientific Committee for Dense Magnetized Plasmas,
http://www.icdmp.pl/isc-dmp
Hery 23, P.O. Box 49, 00-908 Warsaw, Poland

Abstract

This paper describes in detail the construction of a local right-handed, orthogonal curvilinear coordinate system whose axes are along the local tangent, the local azimuth and the local normal of an analytically defined 3D surface of rotation, whose shape mimics the shape of a plasma focus current sheath. Expressions for various differential operators are derived in a tutorial format for the benefit of young researchers and non-specialists. Physical problems expressed in this coordinate system would benefit from the natural symmetry properties of the plasma focus sheath. For example, the normal component of current density is zero and the velocity has mainly the normal component. This paper is meant to serve as a readily available reference in the hope that it would be found useful.

I. Introduction:

The Gratton-Vargas (GV) model[1] of the Dense Plasma Focus (DPF) [2,3] is an analytical kinematic model that attempts a description of some aspects of the plasma focus phenomenon in terms the properties of the solutions of a Jacobi-type partial differential equation based on the snowplow hypothesis. These solutions form a family of 3-dimensional surfaces of rotation that depend only on the scaled geometry of the plasma focus and are independent of the physical size of the device and its operating parameters such as capacitor bank capacity, inductance, resistance, voltage and gas pressure[4-6]. These surfaces depend on a dimensionless time-like parameter $\tau$ that has a dual meaning. On the one hand, as a time-like independent variable of the Jacobi equation, it serves as a parameter describing a family of surfaces. In this role, it is related to the scaled geometry of the plasma focus and mimics the propagation of the plasma sheath in dimensionless coordinate space with respect to dimensionless time. On the other hand, it is related to the real time in terms of the current that has flown in the circuit in that time. This dual role allows a decoupling between the shape and propagation of the solutions in dimensionless parameter space on the one hand and the operating parameters of the plasma focus on the other hand. The details of the model and its derivation [4-6] can be found elsewhere.

The analytically defined 3-D surfaces that mimic the shape and propagation of the plasma focus current sheath allow the possibility of constructing a local right-handed, curvilinear, orthogonal coordinate system whose axes are along the local tangent, along the azimuth and along the local normal to the surface [7] for a classical Mather type plasma focus. Some features of plasma focus physics may have a simpler representation when expressed in coordinates that follow the native geometry of the plasma focus sheath. An extension of the RGV model to arbitrary geometry is under development [6]. The present paper revisits the construction of this local coordinate system in the revised formalism and presents some additional results. Its purpose it to provide intermediate steps of derivations in a tutorial format so that they can be examined by interested readers for their own satisfaction, thereby facilitating discovery of errors, if any. The utility of this formalism is illustrated by applying it to a problem.

II. Derivation of basic relations

Derivation of the Gratton-Vargas partial differential equation and its solution is discussed in detail elsewhere [6] and those results and nomenclature will be used below without repetition. The only relevant aspect of nomenclature is that the cylindrical coordinates are scaled to a scale length 'a' (that is the anode radius in the case of a classical Mather type device) and are then denoted with an overtilde: $(\tilde{r},\tilde{z})$. The unit vectors are written with a caret over the coordinate symbol: $(\hat{r},\hat{z})$

The general scheme of the coordinate system can be understood in terms of the fact that the family of characteristic curves of the GV equation in $(\tilde{r},\tilde{z})$, labeled by a parameter N and the family of its integral curves, labeled by parameter $\tau$ form an orthogonal 2-D system of curves. The integral curves can be revolved about the z axis to generate 3-D surfaces of rotation, called GV surfaces. They have an intrinsic symmetry with respect to the azimuthal coordinate θ. These surfaces are doubly curved: their intersections with the z=constant planes are circles while their intersection with θ=constant planes is a curve in $(\tilde{r},\tilde{z})$ with one end intersecting with the anode profile and the other end terminating at the cathode. The main purpose of constructing the local coordinate system is to understand effects due to this double curvature.

A local coordinate system $(\zeta,\theta,\xi)$ can be constructed with unit vector $\hat{\zeta}$ along the tangent, a unit vector $\hat{\theta}$ along the azimuth and a unit vector along the normal to GV surface defined as $\hat{\xi} \equiv \hat{\zeta} \times \hat{\theta}$ as described below. It is important to note that the family of characteristics labeled by parameter N and the family of the GV surfaces of rotation labeled by parameter $\tau$ do not map the entire $(r,\theta,z)$ space. However, nothing prevents construction of an orthogonal coordinate system with its origin on a chosen point of intersection between one characteristic curve and one integral surface and whose axes are aligned with the two curves. This coordinate system would cover the entire infinite space. A similar coordinate system can be constructed at another such intersection and that too would cover the entire infinite space. Physical laws may be expressed in any such coordinate system and be equally valid. This allows investigation of certain facets of plasma focus physics making use of the symmetry inherent in the native geometry of the plasma focus sheath. For example, it may be a good working hypothesis to assume that the current flows only in the tangential direction or that significant variation of density and temperature occurs only in the direction normal to the sheath.

The physical application of this frame of reference is intended only in a limited region near the origin of the coordinate system, even though it theoretically represents the entire infinite space. In this local coordinate system, the coordinates of arbitrary points are expressed as the numbers $(\zeta,\theta,\xi)$ measured along the tangent to the GV surface, along the azimuthal circle and in the direction perpendicular to the GV surface, all distances being scaled with a scale length a, which is the anode radius for a classical Mather plasma focus. Accordingly, spatial derivatives in the local coordinate system must be divided by a in physical applications.

The curve normal to GV surface is the characteristic curve labeled by parameter N and given by [4,6]

$$\frac{\tilde{z}}{N} + s\text{ArcCosh}\left(\frac{\tilde{r}}{|N|}\right) = C_1 = \text{Constant} \tag{1}$$

where the procedure for determination of constant $C_1$ for an arbitrary geometry is described in detail elsewhere [6]. N is an invariant of the GV partial differential equation, that is determined from the boundary conditions. The slope of the tangent to this curve is

$$\left(\frac{d\tilde{z}}{d\tilde{r}}\right)_1 = -s\frac{N}{\sqrt{\tilde{r}^2 - N^2}}; \quad s = \pm 1 \tag{2}$$

The slope of a line perpendicular to this is

$$\left(\frac{d\tilde{z}}{d\tilde{r}}\right)_2 = -\frac{1}{\left(d\tilde{z}/d\tilde{r}\right)_1} = s\frac{\sqrt{\tilde{r}^2 - N^2}}{N} \tag{3}$$

This is the slope of the tangent to the GV surface.

The equation of a line element along the tangent to GV surface will be

$$dT = d\tilde{z}\hat{z} + d\tilde{r}\hat{r} = \left(\frac{d\tilde{z}}{d\tilde{r}}\right)_2 d\tilde{r}\hat{z} + d\tilde{r}\hat{r}$$
$$= \left\{s\frac{\sqrt{\tilde{r}^2 - N^2}}{N}\hat{z} + \hat{r}\right\}d\tilde{r} \tag{4}$$

The unit vector along tangent to GV surface will be

$$\hat{\zeta} = \frac{dT}{|dT|} = \hat{z}s\tilde{r}^{-1}\sqrt{\tilde{r}^2 - N^2} + N\tilde{r}^{-1}\hat{r} \tag{5}$$

From orthogonality of unit vectors,

$$\hat{\xi} \equiv \hat{\zeta} \times \hat{\theta}$$
$$= \left(\hat{z} \times \hat{\theta}s\tilde{r}^{-1}\sqrt{\tilde{r}^2 - N^2} + N\tilde{r}^{-1}\hat{r} \times \hat{\theta}\right) \tag{6}$$
$$= \left(-\hat{r}s\tilde{r}^{-1}\sqrt{\tilde{r}^2 - N^2} + N\tilde{r}^{-1}\hat{z}\right)$$

Relations (5) and (6) can be inverted

$$\hat{r} = \hat{\zeta}N\tilde{r}^{-1} - \hat{\xi}s\tilde{r}^{-1}\sqrt{\tilde{r}^2 - N^2}$$
$$\hat{z} = \hat{\zeta}s\tilde{r}^{-1}\sqrt{\tilde{r}^2 - N^2} + \hat{\xi}N\tilde{r}^{-1} \tag{7}$$

Equate the line element in the two coordinate systems

$$\hat{r}d\tilde{r} + \hat{z}d\tilde{z} = \hat{\xi}d\xi + \hat{\zeta}d\zeta$$
$$= \left(-\hat{r}s\tilde{r}^{-1}\sqrt{\tilde{r}^2 - N^2} + N\tilde{r}^{-1}\hat{z}\right)d\xi + \left(\hat{z}s\tilde{r}^{-1}\sqrt{\tilde{r}^2 - N^2} + N\tilde{r}^{-1}\hat{r}\right)d\zeta \qquad (8)$$
$$= \hat{r}\left(-d\xi s\tilde{r}^{-1}\sqrt{\tilde{r}^2 - N^2} + d\zeta N\tilde{r}^{-1}\right) + \hat{z}\left(d\xi N\tilde{r}^{-1} + d\zeta s\tilde{r}^{-1}\sqrt{\tilde{r}^2 - N^2}\right)$$

Equating r and z components

$$\begin{aligned} d\tilde{r} &= -d\xi s\tilde{r}^{-1}\sqrt{\tilde{r}^2 - N^2} + d\zeta N\tilde{r}^{-1} \\ d\tilde{z} &= d\xi N\tilde{r}^{-1} + d\zeta s\tilde{r}^{-1}\sqrt{\tilde{r}^2 - N^2} \end{aligned} \quad ; \quad \begin{aligned} \partial_\xi \tilde{r} &= -s\tilde{r}^{-1}\sqrt{\tilde{r}^2 - N^2} \\ \partial_\zeta \tilde{r} &= N\tilde{r}^{-1} \end{aligned} \qquad (9)$$

Solve for $d\xi$, $d\zeta$

$$\begin{aligned} d\xi &= -d\tilde{r}s\tilde{r}^{-1}\sqrt{\tilde{r}^2 - N^2} + d\tilde{z}N\tilde{r}^{-1} \\ d\zeta &= N\tilde{r}^{-1}d\tilde{r} + s\tilde{r}^{-1}\sqrt{\tilde{r}^2 - N^2}d\tilde{z} \end{aligned} \qquad (10)$$

Using (9), the unit vectors can also be written as

$$\begin{aligned} \hat{\zeta} &= -\hat{z}\partial_\xi \tilde{r} + \hat{r}\partial_\zeta \tilde{r} \\ \hat{\xi} &= \hat{r}\partial_\xi \tilde{r} + \hat{z}\partial_\zeta \tilde{r} \end{aligned} \qquad (11)$$

$$\begin{aligned} \hat{r} &= \hat{\zeta}\partial_\zeta \tilde{r} + \hat{\xi}\partial_\xi \tilde{r} \\ \hat{z} &= -\partial_\xi \tilde{r}\hat{\zeta} + \hat{\xi}\partial_\zeta \tilde{r} \end{aligned} \qquad (12)$$

Mutual orthogonality of the unit vectors follows from the definition of $\hat{\xi}$:

$$\hat{\xi} \equiv \hat{\zeta} \times \hat{\theta}; \quad \hat{\xi} \times \hat{\zeta} = -\hat{\zeta} \times \left(\hat{\zeta} \times \hat{\theta}\right) = \hat{\theta}; \quad \hat{\theta} \times \hat{\xi} = \hat{\theta} \times \left(\hat{\zeta} \times \hat{\theta}\right) = \hat{\zeta} \qquad (13)$$

These relations can be used to calculate the derivatives of the unit vectors in the new system using the well-known results for the cylindrical coordinate system

$$\begin{aligned} \partial_{\tilde{r}}\hat{r} &= 0; & \partial_{\tilde{r}}\hat{\theta} &= 0; & \partial_{\tilde{r}}\hat{z} &= 0 \\ \partial_\theta \hat{r} &= \hat{\theta}; & \partial_\theta \hat{\theta} &= -\hat{r}; & \partial_\theta \hat{z} &= 0 \\ \partial_{\tilde{z}}\hat{r} &= 0; & \partial_{\tilde{z}}\hat{\theta} &= 0; & \partial_{\tilde{z}}\hat{z} &= 0 \end{aligned} \qquad (14)$$

Then, the following results are readily seen

$$\partial_\xi \hat{r} = \partial_\xi \tilde{r} \cancel{\partial_{\tilde{r}} \hat{r}} + \cancel{\partial_\xi \theta} \partial_\theta \hat{r} + \partial_\xi \tilde{z} \cancel{\partial_{\tilde{z}} \hat{r}} = 0;$$

$$\partial_\xi \hat{\theta} = \partial_\xi \tilde{r} \cancel{\partial_{\tilde{r}} \hat{\theta}} + \cancel{\partial_\xi \theta} \partial_\theta \hat{\theta} + \partial_\xi \tilde{z} \cancel{\partial_{\tilde{z}} \hat{\theta}} = 0 \quad (15)$$

$$\partial_\xi \hat{z} = \partial_\xi \tilde{r} \cancel{\partial_{\tilde{r}} \hat{z}} + +\partial_\xi \theta \cancel{\partial_{\theta} \hat{z}} + \partial_\xi \tilde{z} \cancel{\partial_{\tilde{z}} \hat{z}} = 0$$

$$\partial_\zeta \hat{r} = \partial_\zeta \tilde{r} \cancel{\partial_{\tilde{r}} \hat{r}} + \cancel{\partial_\zeta \theta} \partial_\theta \hat{r} + \partial_\zeta \tilde{z} \cancel{\partial_{\tilde{z}} \hat{r}} = 0;$$

$$\partial_\zeta \hat{\theta} = \partial_\zeta \tilde{r} \cancel{\partial_{\tilde{r}} \hat{\theta}} + \cancel{\partial_\zeta \theta} \partial_\theta \hat{\theta} + \partial_\zeta \tilde{z} \cancel{\partial_{\tilde{z}} \hat{\theta}} = 0 \quad (16)$$

$$\partial_\zeta \hat{z} = \partial_\zeta \tilde{r} \cancel{\partial_{\tilde{r}} \hat{z}} + +\partial_\zeta \theta \cancel{\partial_{\theta} \hat{z}} + \partial_\zeta \tilde{z} \cancel{\partial_{\tilde{z}} \hat{z}} = 0$$

Then,

$$\begin{aligned}
\partial_\zeta \hat{\zeta} &= -\partial_\xi \tilde{r} \cancel{\partial_\zeta (\hat{z})} - \hat{z}\partial_\zeta(\partial_\xi \tilde{r}) + \partial_\zeta \tilde{r} \cancel{\partial_\zeta (\hat{r})} + \hat{r}\partial_\zeta(\partial_\zeta \tilde{r}) \\
&= -\hat{z}\partial_\zeta(\partial_\xi \tilde{r}) + \hat{r}\partial_\zeta(\partial_\zeta \tilde{r}) \\
&= -\left(-\partial_\xi \tilde{r}\hat{\zeta} + \hat{\xi}\partial_\zeta \tilde{r}\right)\partial_\zeta(\partial_\xi \tilde{r}) + \left(\hat{\zeta}\partial_\zeta \tilde{r} + \hat{\xi}\partial_\xi \tilde{r}\right)\partial_\zeta(\partial_\zeta \tilde{r}) \\
&= +\hat{\zeta}\left(\frac{\partial \tilde{r}}{\partial \xi}\frac{\partial^2 \tilde{r}}{\partial \zeta \partial \xi} + \frac{\partial \tilde{r}}{\partial \zeta}\frac{\partial^2 \tilde{r}}{\partial \zeta^2}\right) + \hat{\xi}\left(-\frac{\partial \tilde{r}}{\partial \zeta}\frac{\partial^2 \tilde{r}}{\partial \zeta \partial \xi} + \frac{\partial^2 \tilde{r}}{\partial \zeta^2}\frac{\partial \tilde{r}}{\partial \xi}\right)
\end{aligned} \quad (17)$$

$$\partial_\zeta \hat{\theta} = \partial_\zeta \tilde{r} \cancel{\partial_{\tilde{r}} \hat{\theta}} + \cancel{\partial_\zeta \theta} \partial_\theta \hat{\theta} + \partial_\zeta \tilde{z} \cancel{\partial_{\tilde{z}} \hat{\theta}} = 0 \quad (18)$$

$$\begin{aligned}
\partial_\zeta \hat{\xi} &= \partial_\zeta\left(\hat{r}\partial_\xi \tilde{r} + \hat{z}\partial_\zeta \tilde{r}\right) \\
&= \partial_\xi \tilde{r} \cancel{\partial_\zeta(\hat{r})} + \hat{r}\partial_\zeta(\partial_\xi \tilde{r}) + \partial_\zeta \tilde{r} \cancel{\partial_\zeta(\hat{z})} + \hat{z}\partial_\zeta(\partial_\zeta \tilde{r}) \\
&= +\left(\hat{\zeta}\partial_\zeta \tilde{r} + \hat{\xi}\partial_\xi \tilde{r}\right)\partial_\zeta(\partial_\xi \tilde{r}) + \left(-\partial_\xi \tilde{r}\hat{\zeta} + \hat{\xi}\partial_\zeta \tilde{r}\right)\partial_\zeta(\partial_\zeta \tilde{r}) \\
&= +\hat{\zeta}\left(\frac{\partial \tilde{r}}{\partial \zeta}\frac{\partial^2 \tilde{r}}{\partial \zeta \partial \xi} - \frac{\partial \tilde{r}}{\partial \xi}\frac{\partial^2 \tilde{r}}{\partial \zeta^2}\right) + \hat{\xi}\left(\frac{\partial \tilde{r}}{\partial \xi}\frac{\partial^2 \tilde{r}}{\partial \zeta \partial \xi} + \frac{\partial \tilde{r}}{\partial \zeta}\frac{\partial^2 \tilde{r}}{\partial \zeta^2}\right)
\end{aligned} \quad (19)$$

$$\begin{aligned}
\partial_\theta \hat{\zeta} &= \partial_\theta\left(-\hat{z}\partial_\xi \tilde{r} + \hat{r}\partial_\zeta \tilde{r}\right) \\
&= -\partial_\xi \tilde{r} \cancel{\partial_\theta(\hat{z})} - \hat{z}\cancel{\partial_\theta(\partial_\xi \tilde{r})} + \partial_\zeta \tilde{r}\partial_\theta(\hat{r}) + \hat{r}\cancel{\partial_\theta(\partial_\zeta \tilde{r})} \\
&= +\hat{\theta}\partial_\zeta \tilde{r}
\end{aligned} \quad (20)$$

$$\partial_\theta \hat{\theta} = -\hat{r} = -\hat{\zeta}\partial_\zeta \tilde{r} - \hat{\xi}\partial_\xi \tilde{r} \tag{21}$$

$$\begin{aligned}\partial_\theta \hat{\xi} &= \partial_\theta \left(\hat{r}\partial_\xi \tilde{r} + \hat{z}\partial_\zeta \tilde{r}\right) \\ &= \partial_\xi \tilde{r}\partial_\theta (\hat{r}) + \hat{r}\cancel{\partial_\theta(\partial_\xi \tilde{r})} + \partial_\zeta \tilde{r}\cancel{\partial_\theta(\hat{z})} + \hat{z}\cancel{\partial_\theta(\partial_\zeta \tilde{r})} \\ &= \partial_\xi \tilde{r}\partial_\theta(\hat{r}) = \hat{\theta}\partial_\xi \tilde{r}\end{aligned} \tag{22}$$

$$\begin{aligned}\partial_\theta \hat{\xi} &= \partial_\theta \left(\hat{r}\partial_\xi \tilde{r} + \hat{z}\partial_\zeta \tilde{r}\right) \\ &= \partial_\xi \tilde{r}\partial_\theta (\hat{r}) + \hat{r}\cancel{\partial_\theta(\partial_\xi \tilde{r})} + \partial_\zeta \tilde{r}\cancel{\partial_\theta(\hat{z})} + \hat{z}\cancel{\partial_\theta(\partial_\zeta \tilde{r})} \\ &= \hat{\theta}\partial_\xi \tilde{r}\end{aligned} \tag{23}$$

$$\begin{aligned}\partial_\xi \hat{\zeta} &= \partial_\xi \left(-\hat{z}\partial_\xi \tilde{r} + \hat{r}\partial_\zeta \tilde{r}\right) \\ &= -(\partial_\xi \tilde{r})\cancel{\partial_\xi(\hat{z})} - \hat{z}\partial_\xi(\partial_\xi \tilde{r}) + \partial_\zeta \tilde{r}\cancel{\partial_\xi(\hat{r})} + \hat{r}\partial_\xi(\partial_\zeta \tilde{r}) \\ &= -\left(-\hat{\zeta}\partial_\xi \tilde{r} + \hat{\xi}\partial_\zeta \tilde{r}\right)\partial_\xi(\partial_\xi \tilde{r}) + \left(\hat{\zeta}\partial_\zeta \tilde{r} + \hat{\xi}\partial_\xi \tilde{r}\right)\partial_\xi(\partial_\zeta \tilde{r}) \\ &= +\hat{\zeta}\left(\frac{\partial^2 \tilde{r}}{\partial \xi^2}\frac{\partial \tilde{r}}{\partial \xi} + \frac{\partial^2 \tilde{r}}{\partial \xi \partial \zeta}\frac{\partial \tilde{r}}{\partial \zeta}\right) + \hat{\xi}\left(\frac{\partial^2 \tilde{r}}{\partial \xi \partial \zeta}\frac{\partial \tilde{r}}{\partial \xi} - \frac{\partial^2 \tilde{r}}{\partial \xi^2}\frac{\partial \tilde{r}}{\partial \zeta}\right)\end{aligned} \tag{24}$$

$$\partial_\xi \hat{\theta} = \partial_\xi \tilde{r}\cancel{\partial_{\hat{r}}\hat{\theta}} + \cancel{\partial_\xi \theta \partial_\theta \hat{\theta}} + \partial_\xi \tilde{z}\cancel{\partial_{\hat{z}}\hat{\theta}} = 0 \tag{25}$$

$$\begin{aligned}\partial_\xi \hat{\xi} &= \partial_\xi \left(\hat{r}\partial_\xi \tilde{r} + \hat{z}\partial_\zeta \tilde{r}\right) \\ &= \partial_\xi \tilde{r}\cancel{\partial_\xi(\hat{r})} + \hat{r}\partial_\xi(\partial_\xi \tilde{r}) + \partial_\zeta \tilde{r}\cancel{\partial_\xi(\hat{z})} + \hat{z}\partial_\xi(\partial_\zeta \tilde{r}) \\ &= \left(\hat{\zeta}\partial_\zeta \tilde{r} + \hat{\xi}\partial_\xi \tilde{r}\right)\partial_\xi(\partial_\xi \tilde{r}) + \left(-\hat{\zeta}\partial_\xi \tilde{r} + \hat{\xi}\partial_\zeta \tilde{r}\right)\partial_\xi(\partial_\zeta \tilde{r}) \\ &= \hat{\zeta}\left(\frac{\partial \tilde{r}}{\partial \zeta}\frac{\partial^2 \tilde{r}}{\partial \xi^2} - \frac{\partial \tilde{r}}{\partial \xi}\frac{\partial^2 \tilde{r}}{\partial \xi \partial \zeta}\right) + \hat{\xi}\left(\frac{\partial^2 \tilde{r}}{\partial \xi^2}\frac{\partial \tilde{r}}{\partial \xi} + \frac{\partial^2 \tilde{r}}{\partial \xi \partial \zeta}\frac{\partial \tilde{r}}{\partial \zeta}\right)\end{aligned} \tag{26}$$

It is seen that the function $\tilde{r}(\xi,\zeta)$ and its derivatives play a central role in differential operators. This function is determined as follows.

In the normal direction,

$$d\zeta = 0 \quad \Rightarrow \quad \frac{-N\tilde{r}^{-1}d\tilde{r}}{s\tilde{r}^{-1}\sqrt{\tilde{r}^2 - N^2}} = d\tilde{z} \tag{27}$$

Substitution in the expression (10) for $d\xi$ gives upon simplification

$$d\xi = -\frac{s\tilde{r}d\tilde{r}}{\sqrt{\tilde{r}^2 - N^2}} \qquad (28)$$

In the normal direction, N is constant. So integration of (28) gives

$$\xi = \xi_0 - s\sqrt{\tilde{r}^2 - N^2} \qquad (29)$$

The constant of integration is chosen from the condition that on a GV surface, $\xi = 0$ by definition. Every point on integral surface of the GV equation can be parametrically represented as

$$\tilde{r}_{GV}(\alpha, N) = |N|\text{Cosh}(\alpha/2);$$
$$\tilde{z}_{GV}(\alpha, N, s) = NC_1(\tilde{Z}, N) - Ns\alpha/2 \qquad (30)$$

where $\alpha$ satisfies the equation

$$F(\alpha) \equiv \text{Sinh}(\alpha) + \alpha = 2\left(C_2(\tilde{Z}, N, s) - \frac{s\tau}{N^2}\right) \qquad (31)$$

where $C_2(\tilde{Z}, N, s)$ is a constant of integration which is determined using the formalism of Ref [6]. Here $\tilde{Z}$ is a parameter that is related to the shape of the anode. The parameter $\alpha(\tilde{Z}, N, s, \tau)$ is determined using the inverse function $F^{-1}$. From (29) and (30)

$$\xi_0 = s|N_0|\text{Sinh}(\alpha_0/2) \qquad (32)$$

where, $N_0$ and $\alpha_0$ refer to the intersection of the characteristic curve and the GV surface that defines the origin of the local coordinate system. This is a constant of the coordinate system.

In the tangential direction, $\xi$ is constant but N is not. From (10)

$$d\xi = 0 \quad \Rightarrow \quad \frac{d\tilde{r}s\tilde{r}^{-1}\sqrt{\tilde{r}^2 - N^2}}{N\tilde{r}^{-1}} = d\tilde{z} \qquad (33)$$

Substitute in expression (10) for $d\zeta$ and simplify to get

$$d\zeta = N^{-1}\tilde{r}d\tilde{r} \qquad (34)$$

From (29),

$$(\xi - \xi_0)^2 + N^2 = \tilde{r}^2 \qquad (35)$$

From (35) and (34)

$$d\zeta = \frac{\tilde{r}d\tilde{r}}{\pm\sqrt{\tilde{r}^2 - (\xi - \xi_0)^2}} \tag{36}$$

Integration gives

$$\zeta = \zeta_0 \pm \sqrt{\tilde{r}^2 - (\xi - \xi_0)^2} \tag{37}$$

The tangential distance must increase with increasing radial distance, indicating that the positive sign must be chosen. At the origin of the local coordinate system,

$$\zeta = \zeta_0 + \sqrt{\tilde{r}_{GV}^2 - \xi_0^2} = 0$$
$$\Rightarrow \zeta_0 = -N_0\sqrt{\text{Cosh}^2(\alpha_0/2) - \text{Sinh}^2(\alpha_0/2)} = -N_0 \tag{38}$$

From (38) and (37)

$$\tilde{r}(\xi,\zeta) = \sqrt{(\zeta - \zeta_0)^2 + (\xi - \xi_0)^2}$$
$$= \sqrt{(\zeta + N_0)^2 + (\xi - sN_0 \sinh(\alpha_0/2))^2} \tag{39}$$

Using (39), the derivatives of unit vectors can be summarized where $\tilde{r}(\xi,\zeta)$ has been abbreviated as $\tilde{r}$.

$$\partial_\zeta \hat{\zeta} = +\hat{\xi}\frac{(\xi - \xi_0)}{\tilde{r}^2} \tag{40}$$

$$\partial_\zeta \hat{\theta} = 0$$

$$\partial_\zeta \hat{\xi} = -\hat{\zeta}\frac{(\xi - \xi_0)}{\tilde{r}^2} \tag{41}$$

$$\partial_\theta \hat{\zeta} = +\hat{\theta}\frac{(\zeta - \zeta_0)}{\tilde{r}} \tag{42}$$

$$\partial_\theta \hat{\theta} = -\hat{\zeta}\frac{(\zeta - \zeta_0)}{\tilde{r}} - \hat{\xi}\frac{(\xi - \xi_0)}{\tilde{r}} \tag{43}$$

$$\partial_\theta \hat{\xi} = \hat{\theta} \frac{(\xi - \xi_0)}{\tilde{r}} \tag{44}$$

$$\partial_\zeta \hat{\zeta} = -\hat{\xi} \frac{(\zeta - \zeta_0)}{\tilde{r}^2} \tag{45}$$

$$\partial_\xi \hat{\theta} = 0 \tag{46}$$

$$\partial_\zeta \hat{\xi} = \hat{\zeta} \frac{(\zeta - \zeta_0)}{\tilde{r}^2} \tag{47}$$

From (9)

$$d\tilde{z} = d\xi \frac{(\zeta - \zeta_0)}{\tilde{r}} - \frac{(\xi - \xi_0)}{\tilde{r}} d\zeta \tag{48}$$

Integration of (48) gives

$$\tilde{z}(\zeta, \xi) = (\zeta - \zeta_0) \arctan\left(\frac{(\xi - \xi_0)}{\tilde{r}}\right) - (\xi - \xi_0) \arctan\left(\frac{(\zeta - \zeta_0)}{\tilde{r}}\right) \tag{49}$$

Consider now the metric for the local coordinate system.

$$ds^2 = d\xi^2 + \tilde{r}^2(\xi, \zeta) d\theta^2 + d\zeta^2 \tag{50}$$

The scale factors are then : $h_1 = 1, h_2 = \tilde{r}(\xi, \zeta), h_3 = 1$.

The differential operators in this local normalized coordinate system are then

$$\operatorname{grad}(f) \equiv \vec{\nabla} f = \hat{\zeta} \frac{\partial f}{\partial \zeta} + \hat{\theta} \frac{1}{\tilde{r}} \frac{\partial f}{\partial \theta} + \hat{\xi} \frac{\partial f}{\partial \xi} \tag{51}$$

$$\operatorname{div}(\vec{F}) \equiv \vec{\nabla} \cdot \vec{F} = \frac{1}{\tilde{r}} \frac{\partial}{\partial \zeta}(\tilde{r} F_\zeta) + \frac{1}{\tilde{r}} \frac{\partial F_\theta}{\partial \theta} + \frac{1}{\tilde{r}} \frac{\partial}{\partial \xi}(\tilde{r} F_\xi) \tag{52}$$

$$\operatorname{curl}(\vec{F}) \equiv \vec{\nabla} \times \vec{F} = \hat{\zeta}\left(\frac{1}{\tilde{r}} \frac{\partial F_\xi}{\partial \theta} - \frac{1}{\tilde{r}} \frac{\partial \tilde{r} F_\theta}{\partial \xi}\right) + \hat{\theta}\left(\frac{\partial F_\zeta}{\partial \xi} - \frac{\partial F_\xi}{\partial \zeta}\right) + \hat{\xi}\left(\frac{1}{\tilde{r}} \frac{\partial \tilde{r} F_\theta}{\partial \zeta} - \frac{1}{\tilde{r}} \frac{\partial F_\zeta}{\partial \theta}\right) \tag{53}$$

Note that $\vec{\nabla} \times \hat{\zeta} = \vec{\nabla} \times \hat{\theta} = \vec{\nabla} \times \hat{z} = 0$ since (53) does not involve differentiation of the unit vectors. Repeated application of (53) can generate expression for $\vec{\nabla} \times \vec{\nabla} \times \vec{F}$.

$$\nabla^2 f = \vec{\nabla} \cdot \left( \vec{\nabla} f \right) = \vec{\nabla} \cdot \left( \hat{\zeta} \frac{\partial f}{\partial \zeta} + \hat{\theta} \frac{1}{\tilde{r}(\xi,\zeta)} \frac{\partial f}{\partial \theta} + \hat{\xi} \frac{\partial f}{\partial \xi} \right)$$
$$= \frac{1}{\tilde{r}(\xi,\zeta)} \frac{\partial}{\partial \zeta} \left( \tilde{r}(\xi,\zeta) \frac{\partial f}{\partial \zeta} \right) + \frac{1}{\tilde{r}^2(\xi,\zeta)} \frac{\partial^2 f}{\partial \theta^2} + \frac{1}{\tilde{r}(\xi,\zeta)} \frac{\partial}{\partial \xi} \left( \tilde{r}(\xi,\zeta) \frac{\partial f}{\partial \xi} \right)$$
(54)

The derivative $\left( \vec{A} \cdot \vec{\nabla} \right) \vec{B}$ requires a special attention since it occurs in convective derivatives and derivatives of unit vectors play a significant role.

$$\left( \vec{A} \cdot \vec{\nabla} \right) \vec{B} = \left( A_\zeta \frac{\partial}{\partial \zeta} + A_\theta \frac{1}{\tilde{r}(\xi,\zeta)} \frac{\partial}{\partial \theta} + A_\xi \frac{\partial}{\partial \xi} \right) \left( \hat{\zeta} B_\zeta + \hat{\theta} B_\theta + \hat{\xi} B_\xi \right)$$
$$= +\hat{\zeta} \left( A_\zeta \frac{\partial B_\zeta}{\partial \zeta} + A_\theta \frac{1}{\tilde{r}} \frac{\partial B_\zeta}{\partial \theta} + A_\xi \frac{\partial B_\zeta}{\partial \xi} + \frac{1}{\tilde{r}^2} \left( A_\xi B_\xi (\zeta - \zeta_0) - A_\zeta B_\xi (\xi - \xi_0) - \hat{\zeta} A_\theta B_\theta (\zeta - \zeta_0) \right) \right)$$
$$+ \hat{\theta} \left( A_\zeta \frac{\partial B_\theta}{\partial \zeta} + A_\theta \frac{1}{\tilde{r}} \frac{\partial B_\theta}{\partial \theta} + A_\xi \frac{\partial B_\theta}{\partial \xi} + \frac{1}{\tilde{r}^2} \left( A_\theta B_\zeta (\zeta - \zeta_0) + A_\theta B_\xi (\xi - \xi_0) \right) \right)$$
$$+ \hat{\xi} \left( A_\zeta \frac{\partial B_\xi}{\partial \zeta} + A_\theta \frac{1}{\tilde{r}} \frac{\partial B_\xi}{\partial \theta} + A_\xi \frac{\partial B_\xi}{\partial \xi} + \frac{1}{\tilde{r}^2} \left( A_\zeta B_\zeta (\xi - \xi_0) - A_\theta B_\theta (\xi - \xi_0) - A_\xi B_\zeta (\zeta - \zeta_0) \right) \right)$$

More complicated derivatives can be evaluated in a similar fashion as and when required.

Since the GV model is based on azimuthal symmetry, in physical applications it would be proper to assume $\partial_\theta = 0$, which would simplify many expressions.